# First-principles study of 3*d* transition metal atom adsorption onto graphene: the role of the extended line defect

*Zhaoyong Guan,*[†, ‡, ξ, *] *Shuang Ni,*[ξ] *Shuanglin Hu*[#]

[†] School of Chemistry and Chemical Engineering, Shandong University, 250100 Jinan, P. R. China

[‡] Department of Physics, Tsinghua University, Beijing 100084, P. R. China

[ξ] Centre for Multidimensional Carbon Materials, Institute for Basic Science, Ulsan, South Korea

[ξ] Research Center of Laser Fusion, China Academy of Engineering Physics. Mianyang, Sichuan 621900, P. R. China

[#] Institute of Nuclear Physics and Chemistry, China Academy of Engineering Physics, Mianyang, Sichuan 621900, P. R. China

**ABSTRACT**

A type of line defect (LD) composed of alternate squares and octagons (4-8) as the basic unit is currently an experimentally available topological defect in graphene lattice, which brings some interesting modification to magnetic and electronic properties of graphene. The transitional metal (TM) adsorb on graphene with line-defect (4-8), and they show interesting and attractive structural, magnetic and electronic properties. For different TMs such as Fe, Co, Mn, Ni and V, the complex systems show different magnetic and electronic properties. The TM atoms can spontaneously adsorb at quadrangular sites, forming an atomic chain along LD on graphene. The most stable configuration is hollow site of regular tangle. The TMs (TM = Co, Fe, Mn, Ni, V) tend to form extended metal lines, showing ferromagnetic (FM) ground state. For Co, Fe, and V atom, the system are half-metal. The spin-α electron is insulating, while spin-β electron is conductive. For Mn and Ni atom, Mn-LD and Ni-LD present spin-polarized metal; For Fe atom, the Fe-LD shows semimetal with Dirac cones. For Fe and V atom, both Fe-LD and V-LD show spin-polarized half-metallic properties. And its spin-α electron is conducting, while spin-β electron is insulating. Different TMs adsorbing on graphene nanoribbon forming same stable configurations of metal lines, show different electronic properties. The adsorption of TMs introduces magnetism and spin-polarization. These metal lines have potential application in spintronic devices, and work as quasi-one-dimensional metallic wire, which may form building blocks for atomic-scale electrons with well-controlled contacts at atomic level.

# 1. INTRODUCTION

In 2004, Geim et. al. successfully isolated graphene by using the micromechanical method.[1] Graphene is 0 eV gap semimetal,[2] with Dirac cone[3-5] and very high carrier mobility.[6-8] A theoretical study indicated that a half-metallic electronic structure could be realized in a zigzag-edged graphene nanoribbon driven by a transverse electric field.[9-10] Motivated by changing chemical potential,[11-13] edge function,[14-15] chemical doping with boron and nitrogen atoms[11, 13, 16] and domains,[17-18] van der Walls structures,[19-21] and topological line-defect[14, 22-24] are used to tune magnetic and electronic properties of low-dimensional materials. As a result, a completely spin-polarized electronic transport is expected in these nanostructures.[11-14] Limited by strong E-field[9-10] and operated accuracy of chemical functions,[15, 25-27] it's hard to realize in the experiment.[9, 11] As result, spin

polarization of graphene nanoribbon has not yet been observed in the experiments.[14]

Another method of realizing spin polarization and half-metallicity is doped with TM,[16, 28-32] such as Cobalt (Co),[33-36] Iron (Fe),[37-39] Manganese (Mn),[32, 40-41] Nickel (Ni)[42-44] and Vanadium (V).[45] Single TM atom adsorbing on either perfect[46] or defective graphene[29] has been widely investigated,[47-48] showing quite attractive geometrical, magnetic and electronic properties.[45, 48-50] Atomic chain adsorbing on surface is also observed in the experiments.[28, 51] But there are few theoretical work to investigate atomic chains, especial metal atomic chains, on defective graphene and graphene ribbons.[30] It is the objective of the present work to research the adsorption of TM chains on graphene and systematically investigate geometrical, magnetic and electronic properties.

In 2018, Zhong et al. have successfully synthesized periodically embedded four- and eight-membered rings with on-surface method.[52] And graphene with LD are semiconductor with reduced band gap.[52-53] In our previous investigation, the electronic and mechanical properties of graphene with LD consist of square and octagon, have been predicted.[53] The LD could introduce defect states in the original forbidden band gap.[14, 53] And the valance band maximum (VBM) and conduction band minimum (CBM), which are usually relative with the chemical active sites are also transferred from the edge sites to the LD atoms.[53-54] When the TM atoms adsorb on graphene, they intend to adsorb at these defective sites for their high chemical active sites.[28-30, 54-55] Salmeron found water could split graphene and intercalates.[56] Cheng has investigated single TM atom adsorbed on graphene.[46-48] Up to now, researchers have just focused on the single TM atom adsorbing on graphene sheet in their work, but the topological structural, magnetic and electronic properties of TM atom adsorbing on graphene are still unknown. So it's urgent to investigate how TM atom adsorbs on graphene with LD, which consists of octagon and square. Besides, both VBM and CBM of graphene embedded with LD, neither consistent of octagon and pentagon (5-8-5)[14, 53] or square and octagon (4-8)[53] localize at the LD sites. What the effects of the adsorption of TM on graphene are still unknown. Furthermore, well-defined atomic structure of nanowire can help to overcome practically one of the challenges of nanoelectronics.[57] This point is urgently needed for development of molecular electronics,[58-61] single-molecule sensors[62-64] and electro-catalytic energy conversion.[65] Although 1D ferromagnetism is hard to realize in the frame of Ising model, but there are other physical mechanism could explain one-dimensional ferromagnism.[66-68]

## 2. COMPUTATIONAL DETAILS

The calculations of graphene nanoribbon embedded with LD (pair of square and octagon), adsorption with TMs are performed by using a numerical radial function basis set DMol$^3$ package,[69-70] which is based on the density functional theory. The Perdew-Burke-Ernzerhof (PBE)[71] is used to describe the exchange and correlated interaction between electrons. The double numerical atomic orbitals augmented by polarization functions (DNP)[72] is adopted as the basis set. The vacuum space long $z$-direction is set as large as 16 Å to avoid the interaction between imaginary images. The real-space global cutoff radius is set to be 5.6 Å. In order to optimize geometry, calculate energy, density of the states (DOS) and band structures, $8\times3\times1, 16\times6\times1, 20\times8\times1$ and 160 Monkhost-Pack kpoints[73] are sample 2D Brillouin zones, respectively. For TM (TM = Co, Fe, Mn, Ni, V) adsorption on LD-embedded graphene sheet, a large $4\times1\times1$ (five carbon hexagons along $\vec{b}$ direction) supercell is adopted to simulate the isolated TM adsorption, while a smaller $2\times1\times1$ (eleven hexagons along $\vec{b}$ direction) supercell is used to calculate the adsorption of "metal chains" on the graphene sheets. The energy and electron density is converged to $1\times10^{-6}$ au (1 Hartree = 27.21 eV). Geometry optimizations are performed until the corresponding values are less than $2\times10^{-3}$ au/Å on the gradient, $5\times10^{-3}$ Å on the displacement, and $1\times10^{-6}$ au on the total energy, respectively. The graphene is adopted as testing systems to check accuracy. The C−C bond length in graphene is calculated to be 1.42 Å, which is consistent with the experimental value.[6] In order to analyze the interaction strength between TM and graphene, the charge partitioning is calculated by the Hirshfeld method.[74]

## 3. RESULTS AND DISCUSSION

**3.1. Geometry of Graphene with LD.** We first optimize the geometry of graphene with LD, which is consist of successive square and octagon. The corresponding optimized geometry is shown in Figure 1. The corresponding C-C bonds of octagon are 1.50, 1.39 and 1.43 Å, respectively. While for the square, the corresponding C-C bond is 1.50 and 1.40 Å, respectively.[53] Compared with perfect graphene, whose C-C bond is 1.42 Å, some C-C bonds are slightly compressed (1.40 and 1.39 Å), while others are enlarged (1.50 Å). More details can be found in the Figure S1. And this is consistent with the previous simulated results.[52] Six adsorption sites have been investigated. H stands for the hollow site, and $H_1$, $H_2$ and $H_3$ present perfect hollow sites of perfect hexagon, octagon and square, respectively. $B_1$ and $B_2$ show the bridge site of perfect hexagonal and defective rings, respectively. $T_1$ indicates the top site of perfect hexagons.

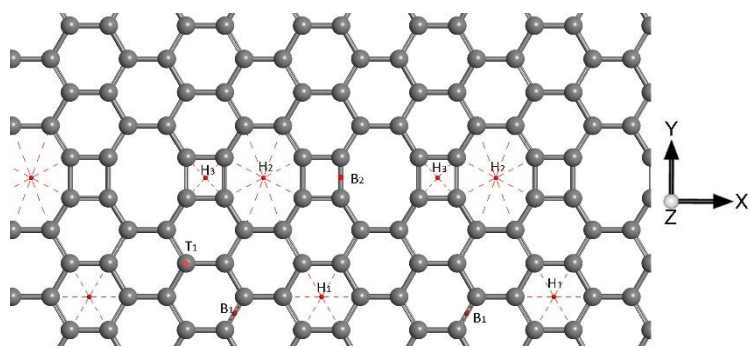

Figure 1. Extended LD made of paired tetragonal and octagonal rings in graphene nanoribbon. $H_1$, $H_2$, $H_3$, $T_1$, $B_1$ and $B_2$ present adsorption sites, hollow, top and bridge sites, respectively. The grey ball presents the carbon atom.

In this work, the adsorption energy ($E_{ad}$) is defined as:

$$E_{ad} = E_{mg} - E_{TM} - E_g$$

The $E_{ad}$ presents the adsorption energy of TM, and $E_{mg}$ presents the energy of the complex TM and graphene. $E_{TM}$ and $E_g$ present the energy of the single TM atom, and energy of the graphene with LD, respectively. $E_{TM}$ is calculated using stable bulk energy of TM divided by number of the atoms.[75]

**3.2 Electronic Structure and Magnetic Properties of TM (TM = Co, Fe, Mn, Ni, V) Adsorption on Graphene.** In this section, the geometry of single TM atom on graphene is investigated. The adsorption configurations including $H_1$, $H_2$, $H_3$, $B_1$, $B_2$, $T_1$ and $T_2$ sites are calculated. The corresponding supercell along lattice $\vec{a}$ and $\vec{b}$ is 17.30 and 13.71 Å, respectively. The distance between metal atom and corresponding image is at least 12 Å to avoid the artificial interaction between metal atom and its image.

**3.2.1 Electronic Structure and Magnetic Properties of Co Atom Adsorption on Graphene.** The $E_{ad}$, the corresponding bond length of the carbon and metal atoms, charge transfer, magnetic moment of system and TM atom are calculated, shown in Table 1. The different adsorption sites $H_1$, $H_2$, $H_3$, $B_1$ correspond to different $E_{ad}$, respectively. Comparing with $H_1$, $H_2$ configurations, the hollow site of square ($H_3$) is the most stable site. And corresponding $E_{ad}$ is 1.62 eV, which is larger than $H_1$ (1.41 eV), $H_2$ (1.12 eV)[45] and $B_1$ (1.04 eV). For $H_3$ site, the corresponding Co-C bond length is 2.00 Å, implying Co atom prefers to stay in the center of square. The $B_1$ site has the smallest $E_{ad}$ (1.04 eV), which means that this site is most unstable in energy. For $H_2$

Table 1. Adsorption energies and structural properties for $H_1$, $H_2$, $H_3$ and $B_1$ sites investigated in this work. The properties listed are $E_{ad}$ (eV), bond length $d_1$, $d_2$, $d_3$, $d_4$ (Å), charge transfer between TM and graphene $\Delta\rho$ ($e$), total magnetic moments MM ($\mu_B$) and TM $MM_{Co}$ ($\mu_B$), respectively.

| Co Sites | $E_{tot}$ (eV) | $E_b$ (eV) | Distance (Å) | | | | $\Delta\rho$ ($e$) | MM ($\mu_B$) | $MM_{Co}$ ($\mu_B$) |
|---|---|---|---|---|---|---|---|---|---|
| | | | $d_1$ | $d_2$ | $d_3$ | $d_4$ | | | |
| $H_2$ | -686.01 | 1.12 | 2.19 | 2.19 | 2.18 | 2.18 | 0.29 | 1.63 | 0.99 |
| $H_3$ | -686.80 | 1.91 | 2.00 | 2.00 | 2.00 | 2.00 | 0.29 | 1.14 | 0.99 |
| $H_1$ | -686.30 | 1.41 | 2.15 | 2.14 | 2.14 | 2.15 | 0.28 | 1.36 | 1.47 |
| $B_1$ | -685.93 | 1.04 | 2.12 | 2.13 | – | – | 0.19 | 2.54 | 2.50 |

and $H_1$ sites, Co atom stays in the center of octagon and hexagon, respectively. And $E_{ad}$ of $H_2$ is little smaller than $H_3$ site, but is bigger than $B_1$ site. It seems that Co atom tends to stay at the hollow site of square, octagon and hexagon. And for hollow site, 0.30 $e$ electron transfers from Co atom to graphene. For bridge site, charge transfer is only 0.19 $e$ electron, which implies weaker interaction between Co atom and graphene. $E_{ad}$, charge transfer, and magnetic properties are also investigated. For $H_2$, $H_3$ and $H_1$, total magnetic moment (MM) is 1.63, 1.14 and 1.36 $\mu_B$, respectively. While for $B_1$, total MM is 2.54 $\mu_B$. Most of magnetic moments come from Co atom, which is confirmed in the spin density of $H_2$, $H_3$, $H_1$ and $B_1$, shown in Figure 3. For $H_1$ configuration, the Co atom contributes 1.72 $\mu_B$ (total MM is 1.47 $\mu_B$), while each nearest nearby carbon atom contributes 0.00 $\mu_B$. Each next nearest nearby carbon atoms contributes 0.03 $\mu_B$ MM, while the other carbon atoms of the square have -0.02 $\mu_B$ MM. For $H_2$ configuration, the Co atom contributes 0.99 $\mu_B$ MM, while other carbon atoms of square contribute -0.02 $\mu_B$ MM. For $H_3$ configuration, the Co atom contributes 1.47 $\mu_B$ MM, while the nearby carbon atoms contribute 0.00, 0.00, -0.00, -0.00, 0.01, 0.01 $\mu_B$ MM, respectively.[46, 48] For $B_1$ configuration, Co atom contributes 2.50 $\mu_B$ MM, while each

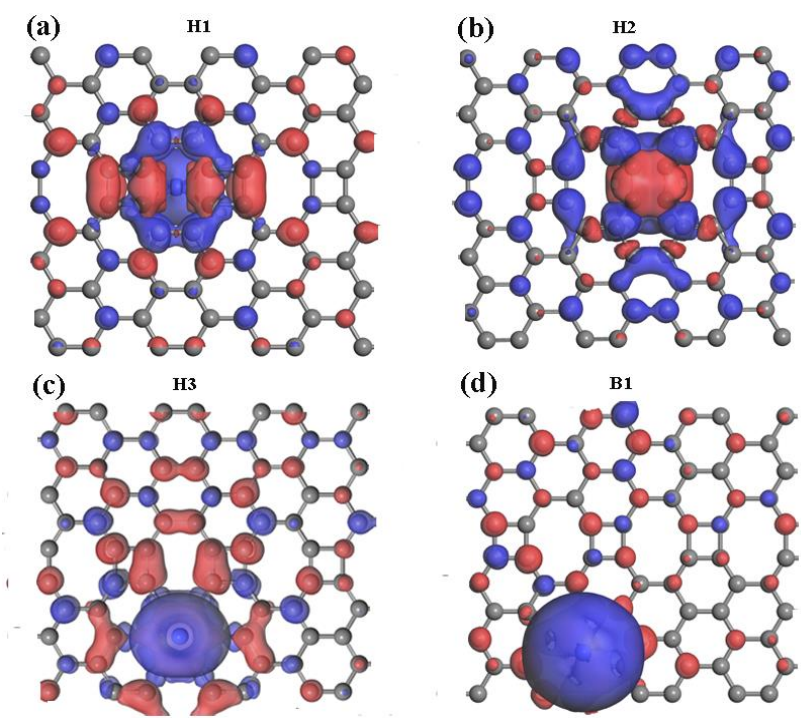

Figure 2. The spin density of Co atom adsorbing on the graphene with different positions. The adsorption sites of (a) hollow site of the octagon ($H_2$), (b) the hollow site of the quadrangle ($H_3$), (c) hollow site of perfect hexagon ($H_1$), and (d) the bridge site ($B_2$) of the hexagon. The isovalue is 0.003 e/Å$^3$.

nearby carbon atoms contributes 0.01 $\mu_B$ MM. From Figure 2, it can conclude that magnetic moment mainly localizes at the Co atom, and it quickly decrease when far away from Co atom.[46, 48]

In the above section, it finds that $H_2$ site is the most stable adsorption site. When the two Co atoms adsorb on graphene, they intend to form a "metallic line".[76] The corresponding bond length of Co-C is 2.01 Å, and the distance between Co atoms is about 4.30 Å, which is larger than corresponding bond length of Co-Co atom (2.51 Å) in the bulk. Two Co atom can ferromagnetically or antiferromagnetically couple with other, and corresponding FM and antiferromagnetic (AFM) spin densities are shown in Figure 3 (a), (b), respectively. The blue and red represents spin-α and spin-β density of electrons, respectively. In Figure 3 (a), both two Co atoms show spin-α density (they have the same color), so two Co atoms ferromagnetically couple with other. Each Co atom has 1.01 $\mu_B$ MM, and Co-LD has 2.00 $\mu_B$ MM. While two Co atoms show different spin-α and spin-β densities, shown in Figure 3 (b), which implies two Co atoms antiferromagnetically couple with other. One Co atom has -1.00 $\mu_B$ MM, while another has 1.00 $\mu_B$ MM, so total MM equals 0.00 $\mu_B$. Each carbon atom of the square has -0.02 $\mu_B$ MM. For carbon atoms of the octagon, they have 0.01 (4 atoms), and -0.02 $\mu_B$ MM (4 atoms), respectively. The MM decreases quickly when it is far away from Co atom. For carbon atoms near the LD, they have as much as 0.01 $\mu_B$ MM, respectively. For carbon atoms only far away from LD, they even have no magnetic moment.[50] Compared with the original graphene with LD, The Co atoms indeed introduces magnetism and spin-polarization.[14, 47-48, 50, 53] The energy difference ($\Delta E$) is defined as energy difference between FM and AFM magnetic configurations, defined as followed:

$$\Delta E = E_{FM} - E_{AFM}$$

$E_{FM}$ and $E_{AFM}$ stand for energy of FM and AFM state, respectively. For Co adsorbing on graphene with LD, the corresponding $\Delta E$ is 0.10 eV, which implies FM ground state. With the mean theory, the Corrie' temperature could be evaluated as followed:

$$T_C = \frac{2}{3k_B} \Delta E$$

Where $\Delta E$ is energy difference between the AFM and FM phase per unit cell, and $k_B$ denotes Boltzmann constant. Here $\Delta E$ can be well approximately chosen to be total energy difference between AFM and FM phases. The corresponding $T_C$ is about 772 K.

Corresponding band structure and DOS is shown in Figure 3 (c) and (d), respectively. From the band structure, we can find that FM ground state is spin-polarized. The spin-α electrons show half-metallic, while and spin-β electrons show semiconductive properties with band gap of 0.30 eV. And this is consistent with analysis of PDOS, shown in Figure 3 (d) and Figure S2. The states near the Fermi-level are mainly contributed by Co atoms, and partially by the carbon atoms (octagon and square), shown in Figure S2. The gap of spin-β electron is 0.30 eV, much larger than the excited energy of electron at room temperature (300 K, 0.026 eV). So the gap is big enough to insure the stability of half-metallicity. The Co atoms tends to form "metallic line", could work as spin filter. When the spin-α and spin-β electrons come across LD adsorption with Co atoms, only spin-α electrons could pass through the graphene nanoribbon, while the spin-β electrons are unable to pass.

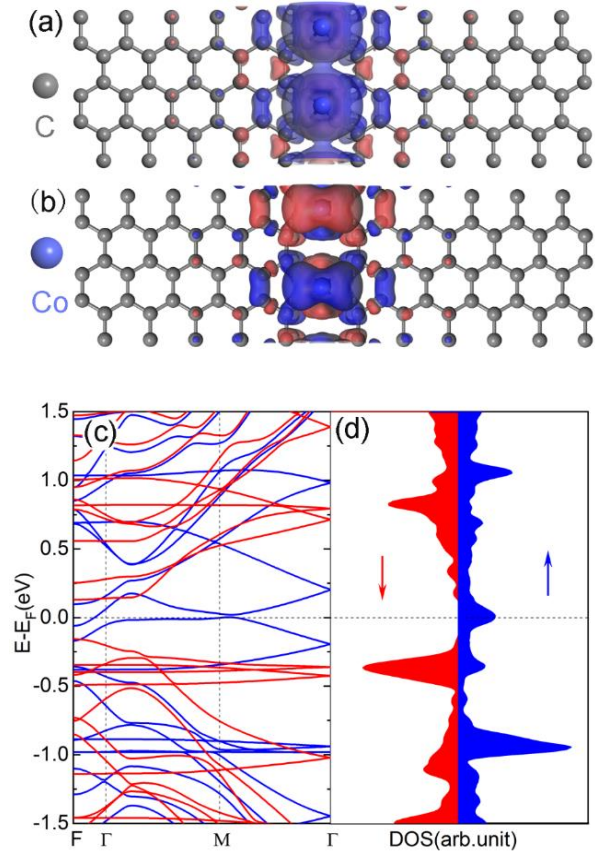

**Figure 3.** (a-d) Spin density, spin-polarized band structure and PDOS of Co-LD. Spin density of (a) FM and (b) AFM configuration of Co-LD. Gray, violet balls present C and Co atoms, respectively. (c) Band structure and (d) DOS of Co-LD. The blue and red lines represent spin-α and spin-β electrons, respectively. Here, the isovalue is set 0.003 e/Å$^3$. The grey, blue balls represent carbon, and cobalt atoms, respectively.

**3.2.2 Electronic Structure and Magnetic Properties of Fe Atom Adsorption on Graphene.** In this section, we investigate structural, magnetic and electronic properties of Fe-LD. First, single Fe atom adsorbing on graphene is studied, and shown in Table 2. From Table 2, it can be found that the most stable adsorption site is $H_3$. The corresponding $E_{ad}$ is about 1.42 eV, with corresponding Fe-C bond is 2.05Å, meaning Fe atom is in the center of the square. While for $H_2$, $H_1$, and $B_1$ sites, the corresponding $E_{ad}$ are 0.90, 1.04, and 0.03 eV, respectively. For $H_1$ configuration, $E_{ad}$ is similar with Fe atom adsorption on perfect graphene, whose $E_{ad}$ is 1.02 eV.[46, 48] The Fe-C bond of $H_3$ site is 2.32 and 2.72 Å, respectively, which is bigger than the later of 2.08 Å.[47] The $E_{ad}$ of most stable $H_3$ site is 0.40 eV bigger than that of $H_1$ site. The $H_2$ site (hollow site of octagon) has 0.10 eV smaller than $H_1$ site in energy. While $B_1$ site is the most unstable site in the considering four sites, it only has 0.03 eV, which means that Fe atom adsorbing at this site is easily shift for the smaller migration barrier.[45] And the corresponding Fe-C bond is 2.35 Å, which is bigger than that $H_1$ site of 2.12 Å.

When the TM atom adsorbs on the graphene, it usually follows the charge transfer. For the $H_2$ site, the Fe atom loses 0.35 $e$ electron. While for $H_2$ and $H_1$ site, the corresponding charge transfer is 0.26 and 0.28 $e$ electron, respectively. For the most unstable $B_1$ site, it is only 0.17 $e$ electron, which means that charge transfer between Fe atom and graphene could be neglected. The charge transfer and $E_{ad}$ show the same trend. Only part of electrons transfer from Fe atom to graphene. For $H_3$ site, it has 2.00 $\mu_B$, and Fe atom contributes 2.45 $\mu_B$ MM, whose spin density is shown in Figure 4 (c). Each carbon atoms of the square has -0.03 $\mu_B$ MM. While the carbon atoms are far away from LD, they have 0.01, -0.01 $\mu_B$ MM, respectively. For carbon atoms far away from LD, there is no magnetic moment distribution. For $H_2$ site, the whole system has 4.02 $\mu_B$ MM, while Fe atom contribute 3.54 $\mu_B$ MM. Each carbon atom of octagon contributes 0.02 and 0.04 $\mu_B$ MM, respectively. For $H_1$ site, the Fe atom has 2.29 $\mu_B$ MM, and whole system has 2.17 $\mu_B$ MM. Each hexagonal carbon atom has -0.02 $\mu_B$ MM, shown in Figure 4 (b). For $B_1$ site, the Fe atom has 2.17 $\mu_B$ MM, and whole system has 4.36 $\mu_B$ MM. Each carbon atom connected to Fe atom has 0.02 $\mu_B$ MM, shown in Figure 4 (d). In a word, it can be found that magnetic moment mainly localizes at the Fe and near C atoms.[48]

Table 2. Adsorption energies and structural properties of $H_1$, $H_2$, $H_3$ and $B_1$ sites investigated in this work. The properties listed are $E_{ad}$ (eV), bond length $d_1$, $d_2$, $d_3$, $d_4$ (Å), the charge transfer between Fe atom and graphene $\Delta\rho$ (e), total magnetic moments MM ($\mu_B$) and TM (irion) $MM_m$ ($\mu_B$), respectively.

| Fe Site | $E_{tot}$ (eV) | $E_{ad}$ (eV) | Distance (Å) | | | | $\Delta\rho$ (e) | MM ($\mu_B$) | $MM_{Fe}$ ($\mu_B$) |
| --- | --- | --- | --- | --- | --- | --- | --- | --- | --- |
| | | | $d_1$ | $d_2$ | $d_3$ | $d_4$ | | | |
| $H_2$ | -685.49 | 0.90 | 2.32 | 2.32 | 2.72 | 2.72 | 0.26 | 4.02 | 3.54 |
| $H_3$ | -686.06 | 1.42 | 2.05 | 2.05 | 2.05 | 2.05 | 0.35 | 2.02 | 2.45 |
| $H_1$ | -685.63 | 1.04 | 2.12 | 2.12 | 2.12 | 2.12 | 0.28 | 2.29 | 2.17 |
| $B_1$ | -684.99 | 0.03 | 2.35 | 2.35 | – | – | 0.17 | 4.36 | 4.17 |

In this part, the magnetic and electronic properties are systematically investigated, and spin density, spin-polarized band structure and DOS of Fe-LD is shown in Figure 5. The Fe atom prefers to stay at $H_3$ site, so two Fe atoms intends to form a "metal line".[30] The corresponding Fe-C bond is 2.07 Å (2.07 Å×4), which is consistent with the $H_3$ site, shown in Figure 4 (c), and the corresponding Fe atom has 2.40 $\mu_B$ MM. Other Fe-C bond is 2.12 Å, which is consistent with $H_1$ site, and Fe atom has 2.40 $\mu_B$ MM. Two Fe atoms lose about 0.38 $e$ electron, and each Fe atom loses 0.19 $e$ electron, respectively. Two Fe atoms have the same spin, so Fe atoms ferromagnetically couple with each other, shown in Figure 5 (a). The carbon atoms of square have the opposite spin of -0.04 and -0.05 $\mu_B$ MM, respectively. The AFM configuration is shown in Figure 5 (b). And one Fe atom has 2.40 $\mu_B$ MM, while another has -2.40 $\mu_B$ MM. Each carbon atoms bounded with Fe atom (2.40 $\mu_B$) has 0.04 $\mu_B$, other carbon atoms has -0.04 $\mu_B$. The $\Delta E$ defined as energy difference between FM and AFM configuration is about 0.05 eV, and the corresponding Tc is 402 K. The spin-polarized band structure and DOS of the FM ground states are also calculated, shown in Figure 5 (c) and (d), respectively. There is a Dirac cone for spin-α electron above the Fermi-level. It implies that Fe-LD persists original Dirac cone of the perfect graphene, but it reduces the original generacy.[6] Besides that the states near the Fermi level are mainly contributed by spin-α electron, while the contribution from spin-β electrons could be eliminated. The spin-α electron is conductive, while spin-β electron is insulator with band gap of 0.13 eV, which implies Fe-LD is half-metal. The gap of spin-β could insure the electron could not be thermally excited from VBM to CBM at room temperature. The PDOS of Fe-LD is also calculated, and result is shown in Figure S4. From PDOS, we can find that the states near the Fermi-level are mainly contributed by Fe atoms. It implies that bonding states are contributed by Fe atoms. While the first peak above Fermi-level also come from the contribution of Fe atoms, implying anti-bond states is composed of spin-β electrons.

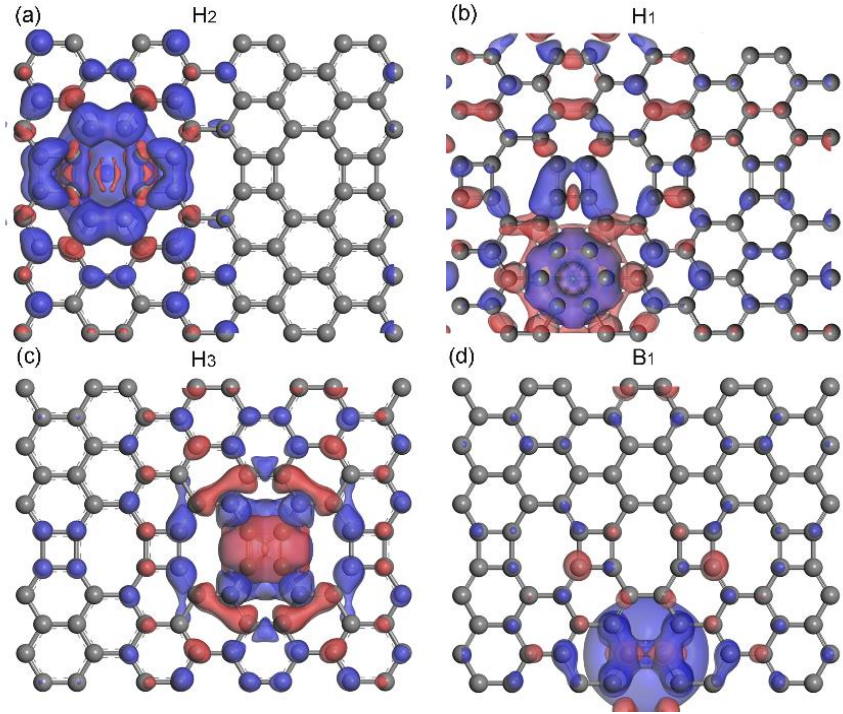

Figure 4. The spin density of the iron adsorption on the graphene nanoribbon. Single iron atom adsorbs on (a) $H_2$, (b) $H_1$, (c) $H_3$ and (d) $B_1$ sites, respectively. The isovalue is set 0.003 e/Å$^3$. The blue and red present spin-α and spin-β electrons, respectively.

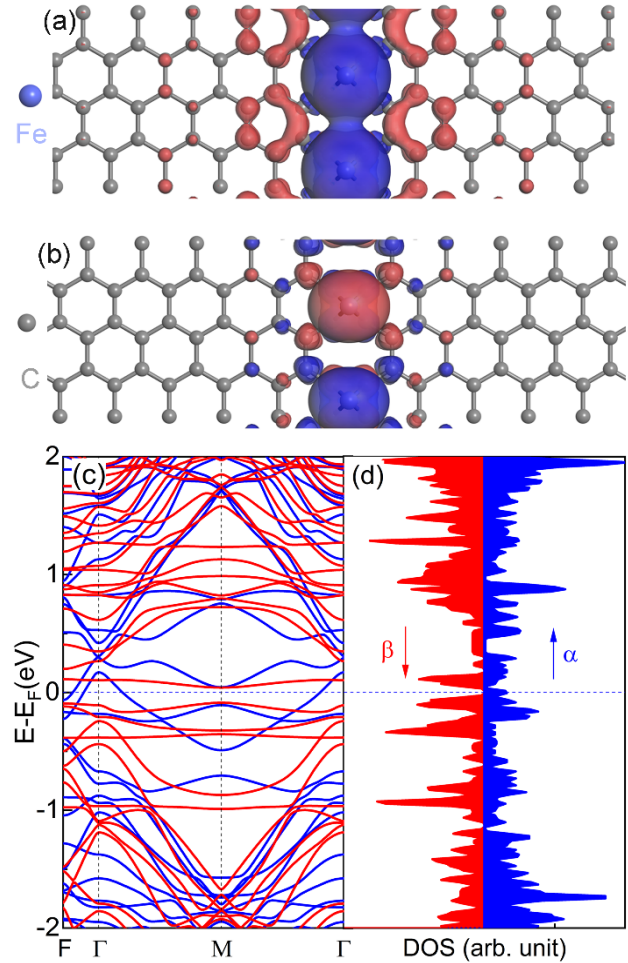

**Figure 5.** (a-d) Spin density, spin-polarized band structure and PDOS of Fe-LD. Spin density of (a) FM and (b) AFM Fe-LD, and blue, grey balls present Fe, carbon atom, respectively. (c) Band structure and (d) PDOS of Fe-LD. The blue and red lines represent spin-α and β electrons, respectively. Here, the isovalue is 0.003 e/Å³. The blue, grey balls represent iron, carbon atoms, respectively.

**3.2.3 Electronic Structure and Magnetic Properties of Mn Atom Adsorption on Graphene.** The structural, magnetic and electronic properties of single Mn atom adsorption on graphene with LD are also systematically investigated. And all results are shown in Table 3. For $H_2$ site, the Mn atom lies in the center of the hexagon, and corresponding Mn-C bond is 2.44 Å. The $E_{ad}$ is 1.16 eV. Mn atom loses 0.31 $e$ electron. For $H_3$ site, Mn-C bond is 2.30 Å, and Mn atoms lies in the center of the square.

**Table 3. Adsorption energies and structural properties of Mn atom for $H_1$, $H_2$, $H_3$ and $B_1$ sites investigated in this work. The properties listed are the $E_{ad}$ (eV), $d_1$, $d_2$, $d_3$, $d_4$ (Å), Δρ (e), MM ($μ_B$) and $MM_{Mn}$ ($μ_B$), respectively.**

| Mn Site | $E_{tot}$ (eV) | $E_{ad}$ (eV) | Distance (Å) $d_1$ | $d_2$ | $d_3$ | $d_4$ | Δρ (e) | MM ($μ_B$) | $MM_{Mn}$ ($μ_B$) |
|---|---|---|---|---|---|---|---|---|---|
| $H_2$ | -684.52 | 1.16 | 2.44 | 2.44 | 2.44 | 2.44 | 0.31 | 5.23 | 5.05 |
| $H_3$ | -684.75 | 1.39 | 2.30 | 2.30 | 2.30 | 2.30 | 0.25 | 5.04 | 5.06 |
| $H_1$ | -684.28 | 0.92 | 2.63 | 2.63 | 2.63 | 2.63 | 0.29 | 5.79 | 5.47 |
| $B_1$ | -683.55 | 0.19 | 2.40 | 2.40 | – | – | 0.26 | 5.67 | 5.42 |

The corresponding $E_{ad}$ is 1.39 eV, which is the biggest value among the considered sites. For $H_1$ site, the corresponding Mn-C bond is 2.63 Å, and $E_{ad}$ is 0.92 eV, which is consistent with the previous result.[47] For $B_1$ site, the corresponding Mn-C bond is

2.40 Å, and $E_{ad}$ is quite small. The Δρ between Mn and graphene is only 0.26 $e$ electron, which is smaller than other configurations.

In the above section, the structural properties of Mn adsorption on graphene are investigated. In this section, the magnetic properties of all kinds of configurations are also listed. For $H_2$ site, Mn atom contributes 5.23 $\mu_B$ MM, while Mn-LD has 5.05 $\mu_B$ MM, each carbon atom of octagon connected to the Mn atom has 0.01 $\mu_B$ MM, while other carbon atoms of octagon have 0.02 $\mu_B$ MM, shown in Figure 6 (b). For $H_3$ configuration, the Mn atom contributes 5.04 $\mu_B$ MM, while the whole system has 5.06 $\mu_B$ MM. Each carbon atoms of the octagon connected to the Mn atom has 0.01 $\mu_B$ MM, while others have 0.02 $\mu_B$ MM, shown in Figure 6 (a). For $H_1$ configuration, the Mn atom lies in the center of the hexagon. And the Mn atom has 5.47 $\mu_B$ MM, while the whole system has 5.79 $\mu_B$ MM. Each carbon atoms of the hexagon connected with the Mn atom contributes 0.03 $\mu_B$ MM. For $B_1$ site, the Mn has 5.42 $\mu_B$ MM, and total MM equals to 5.67 $\mu_B$. Each carbon atoms connected with Mn atom has 0.03 $\mu_B$ MM. And the nearby carbon atoms have 0.03 $\mu_B$ MM. The carbon atom far away from the adsorption site contribute quite small, even could be neglect. For all considered configurations, the spin density mainly localizes at the adsorption sites, and quickly decreases when far away from the adsorption sites. And this point is similar with other TM atoms.

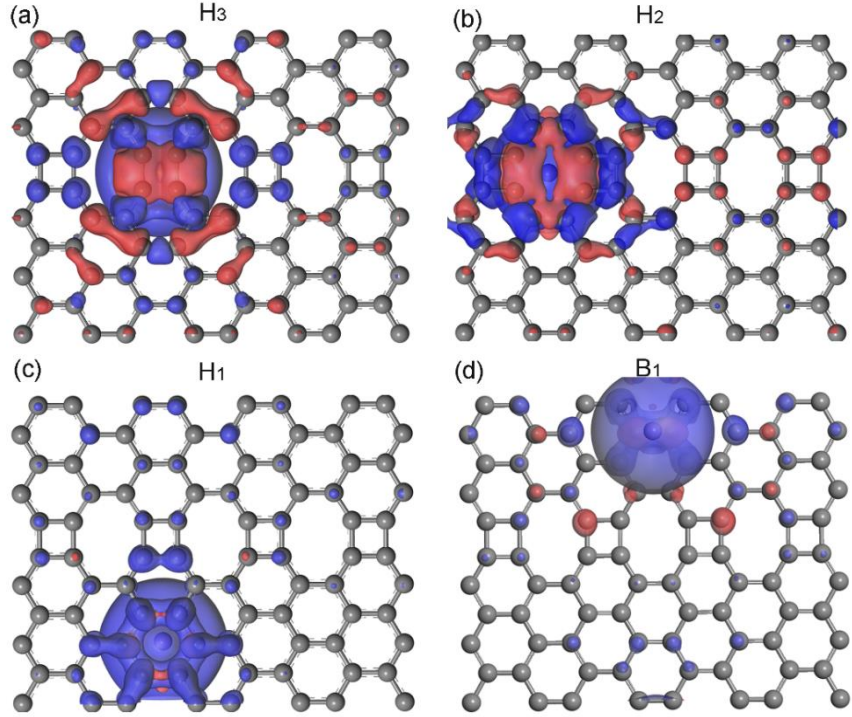

Figure 6. The spin density of Mn atom adsorption on graphene, and the corresponding sites of (a) $H_3$, (b) $H_2$, (c) $H_1$, and (d) $B_1$ is shown. And the corresponding isovalue is set with 0.003 e/Å$^3$.

Among $H_1$, $H_2$, $H_3$ and $B_1$ sites, the most stable adsorption site is $H_3$. So when two Mn atoms adsorption on graphene, two Mn atoms intend to form a "metal line", shown in Figure 7 (a) and (b), respectively. The corresponding Mn-C bond is 2.31 (2.31 Å× 4) and 2.30 Å (2.30 Å×4). Two kinds of magnetic configuration: FM and AFM is considered. And the corresponding spin density is shown in Figure 7 (a), (b), respectively. The energy of FM configuration is -723.82 eV, while the energy of AFM is -723.80 eV. So $\Delta E = 0.02$ eV. And the corresponding $T_C$ is 155 K. For FM configuration, each Mn atom loses 0.24 $e$ electron, which transfers to graphene. Each Mn atom has 5.00 $\mu_B$ MM, and carbon atoms (eight atoms) of square bonded to Mn atom contribute -0.01 $\mu_B$ MM (-0.01 ×8 $\mu_B$). And the whole system has 10.00 $\mu_B$ MM.

For AFM configuration, each Mn atom loses 0.19 $e$ electron. And one Mn atom has

5.00 $\mu_B$ MM, while another has -5.00 $\mu_B$ MM. The carbon atoms of the square connected to Mn atom (5.00 $\mu_B$ MM) has -0.02 $\mu_B$ MM, while carbon atoms connected with Mn atom (-5.00 $\mu_B$ MM) has -0.02 $\mu_B$ MM, shown in Figure 7 (a). The whole system only have 0.00 $\mu_B$, shown in Figure 7 (b). The corresponding $E_{ad}$ is 2.50 eV, which is consistent with the single Mn atom adsorbing on the graphene. The electronic properties of Mn-LDs at FM ground state is also calculated, and shown in Figure 7 (c), (d). From the band structures, we can find that both spin-α and spin-β electrons show contribution at the Fermi-level. So both spin-α and spin-β electrons could come through the Mn-LDs. By the analysis of the PDOS, we can also find that the states near the Fermi-level mainly come the contribution of Mn atoms. Two peaks locating 0.56 and -0.65 eV also come from Mn atoms. Besides, there is several Dirac cones existing in the Mn-LDs. There is a Dirac cone at 0.19 eV above the Fermi-level for the spin-β electrons. There is also Dirac cone at 0.56 eV above the Fermi-level for the spin-β electrons. While for the spin-α electrons, there are two Dirac cones at 0.05 eV above the Fermi-level.

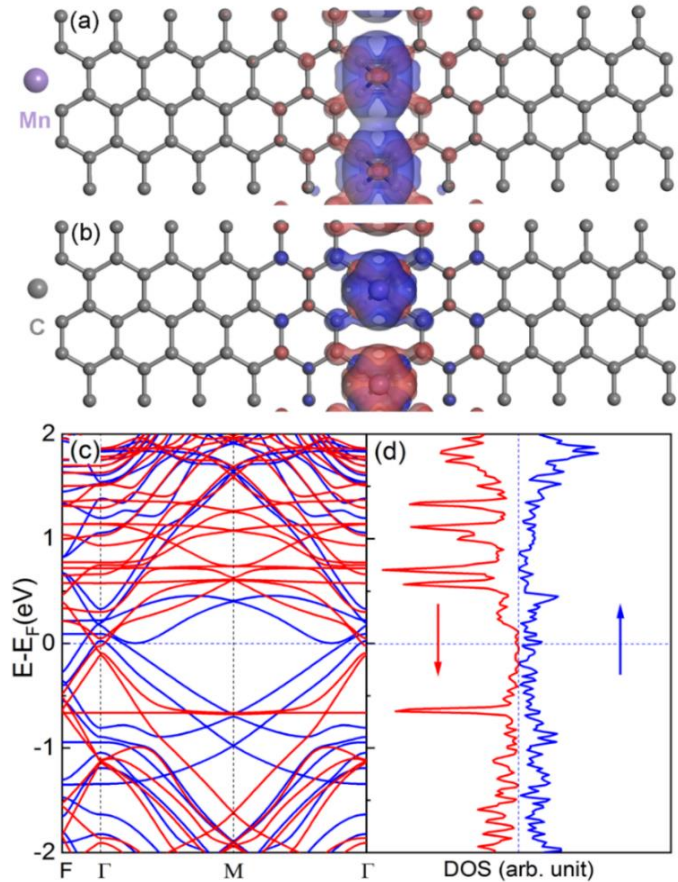

**Figure 7.** The spin density of (a) FM and (b) AFM configuration of Mn-LD. The red and blue lines represent the spin α and β electrons, respectively. (c) Spin-polarized band structure and (d) DOS of Mn-LD. The medium slate blue and grey balls present manganese and carbon atoms, respectively.

**3.2.4 Electronic Structure and Magnetic Properties of Ni Atom Adsorption on Graphene.** When the nano-devices work, they have to be constructed and measured on certain substrates. There is lattice mismatch between all kinds of materials. In order to simply these issues, we use the strain to simulate this situation, and stability of the half-metallicity under the strains is also investigated. Only enlarged strain along lattice $\vec{a}$ is investigated, and spin-polarized band structures under strains of 1% and 2% are also calculated, shown in Figure S3. Though the band gap of spin-β electron is decreased (0.07 eV for 1%, 0.04 eV for 2%), Co-LD still shows half-metallicity. More detail and discussion is provided in Supporting Information.

In this part, structural, magnetic and electronic properties of Ni atom adsorption on graphene are systematically investigated. Different adsorption H$_2$, H$_3$, H$_1$ and B$_1$ sites are investigated, and the results are shown in Table 4. The most stable adsorption site is also H$_3$ site, with the $E_{ad}$ is 2.28 eV, and the corresponding Ni-C bond is 2.03 Å (2.03 Å×4), which is smaller than the Ni-C bond of H$_1$ and H$_2$ site. The Ni atom loses 0.27 $e$ electron. While for H$_2$ site, the corresponding $E_{ad}$ is 1.58 eV, which means that the B$_1$ site is most unstable site in the considering sites. The corresponding Ni-C bond is 2.14 Å (2.14 Å×4). And 0.20 $e$ electron is transferred from Ni atom to the graphene layer. For H$_1$ site, the corresponding $E_{ad}$ is 1.90 eV, and the corresponding Ni-C bond is 2.13 Å (2.13 Å×3) and 2.12 Å (2.12 Å×3), respectively. The graphene gets 0.19 $e$ electron from Ni atom. For the B$_1$ site, the corresponding $E_{ad}$ is 1.72 eV, which is

**Table 4. Adsorption energies and structural properties of Ni atom for the H$_1$, H$_2$, H$_3$ and B$_1$ sites investigated in this project. The properties listed are $E_{ad}$ (eV), bond length d$_1$, d$_2$, d$_3$, d$_4$ (Å), the charge transfer between TM and graphene Δρ (e), total magnetic moments MM (μ$_B$) and Mn MM$_m$ (μ$_B$), respectively.**

| Ni Sites | E$_{tot}$ (eV) | E$_{ad}$ (eV) | d$_1$ | d$_2$ | d$_3$ | d$_4$ | Δρ (e) | MM (μ$_B$) | MM$_{Ni}$ (μ$_B$) |
|---|---|---|---|---|---|---|---|---|---|
| H$_2$ | -686.42 | 1.58 | 2.14 | 2.14 | 2.14 | 2.14 | 0.20 | 0 | 0 |
| H$_3$ | -687.12 | 2.28 | 2.03 | 2.03 | 2.03 | 2.03 | 0.27 | 0 | 0 |
| H$_1$ | -686.74 | 1.90 | 2.13 | 2.13 | 2.12 | 2.12 | 0.19 | 0 | 0 |
| B$_1$ | -686.56 | 1.72 | 1.94 | 1.94 | – | – | 0.21 | 0 | 0 |

smaller than considering H$_1$ and H$_3$ sites. The corresponding Ni-C is 1.94 Å (1.94 Å×2), which is the smallest values in the considered results, shown in Table 4. And graphene gets 0.21 $e$ electron from Ni atom. For all the considered H$_1$, H$_2$, H$_3$ and B$_1$ sites, both

Ni and total MM equals 0.00 $\mu_B$, which is consistent with single Ni atom adsorption on the perfect graphene sheet.[45] Compared with $H_1$ (0.19 $e$), $H_2$ (0.20 $e$), and $B_1$ (0.21 $e$) site, $H_3$ site has the biggest $E_{ad}$, and corresponds to the biggest charge transfer (0.27 $e$).

For the all kinds of adsorption sites, the $H_3$ site has the biggest $E_{ad}$. So the two Ni atom intend to form line, when the two Ni atoms adsorb on the graphene. In this part, the structural, magnetic and electronic properties of the Ni-LD are calculated. The corresponding Ni-C bond is 2.04 Å. Each Ni atom loses about 0.18 $e$ electron, and graphene sheet gets 0.36 $e$ electron from two Ni atoms. Each Ni atom has 0.30 $\mu_B$ MM, and Ni-LD has 1.00 $\mu_B$ MM. The spin density is shown in Figure 8 (a), and each carbon atoms of square and octagon has 0.01 $\mu_B$ MM. The spin density of carbon atoms mainly localize at the defective zones, and it quickly decay when far away from defective areas. The Ni-LDs are at FM ground state, and AFM state is unstable. The $E_{ad}$ of two Ni atoms adsorption at the $H_3$ site is 4.30 eV, which is consistent with the $E_{ad}$ of isolated Ni atom adsorption on graphene. The spin-polarized band structure and density of the state is also investigated, shown in Figure 8. From the band structure, it can be found that Ni-LD is common spin-polarized metal, and both spin-α and spin-β electrons could make contribution to the conduction. The analysis of the DOS is also confirmed above analysis. Both spin-α and spin-β electrons have occupied states at the Fermi-level, shown in Figure 8 (c). Besides that, the states near the Fermi-level mainly come from the contribution of the Ni and defective carbon atoms. And more details could be found in the Figure S8, in the Supplementary Materials. It is also quite interesting that when the single Ni atom adsorbing on the graphene, the whole system show spin-unpolarized ground state. The whole system has no magnetic moment, no matter the Ni atom locates at $H_1$, $H_2$, $H_3$ and $B_1$ sites. The spin-polarization come from two Ni atom ferromagnetically couple with each other. The strain could effectively tune the magnetic and electronic properties of the Ni-LDs.

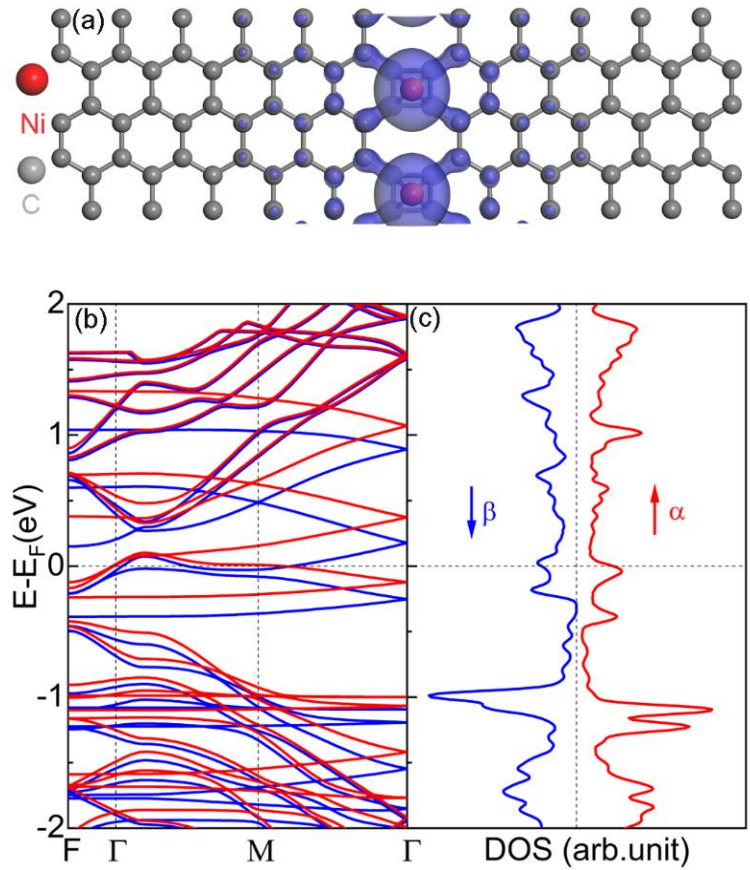

Figure 8. (a-h) The DOS and band structures of Ni-LD. The blue, grey balls present Ni and carbon atoms, respectively. (a-c) spin-polarized of spin density, band structures and PDOS of Ni-LD. Spin density of (a) FM of Ni-LD, (c) Band structure and (d) DOS of Ni-LD is also calculated. The blue and red lines represent the spin-α and spin-β electrons, respectively. Here, the isovalue is set 0.003 e/Å$^3$.

**3.2.5 Electronic Structure and Magnetic Properties of V Atom Adsorption on Graphene.** In the above section, the structural, magnetic and electronic properties of Co, Fe, Mn and Ni atoms are calculated. In the last part, the structural, magnetic and electronic properties of V adsorption on the graphene are also calculated, shown in Table 5. The $E_{ad}$ of $H_2$ site is 2.18 eV, the corresponding V-C bond is 2.21 Å (2.21 Å×4), and there is 0.41 $e$ electron transferring from V atom to the graphene. For the $H_3$ site, the corresponding $E_{ad}$ is -2.52 eV, which is obviously bigger than other sites. The corresponding V-C bond is 2.15 Å (2.15 Å×4), which is smaller than other configurations. While for $H_1$ site, the corresponding V-C bond is 2.31 Å (2.31 Å×3) and 2.32 Å (2.32 Å×3), and the corresponding $E_{ad}$ is 1.82 eV, which is consistent with perfect graphene.[45] For $B_2$ site, the V-C bond is 2.17 Å (2.17 Å×2), 2.16 Å (2.16 Å×2), and the corresponding adsorption energy is 1.75 eV, which is the smallest values in the considering sites, implying this site is less stable than other sites.

**Table 5. Adsorption energies and structural properties of V atom for $H_1$, $H_2$, $H_3$ and $B_1$ sites investigated in this work. The properties listed are $E_{ad}$ (eV), bond length $d_1$, $d_2$, $d_3$, $d_4$ (Å), the charge transfer between TM and graphene Δρ (e), total magnetic moments MM ($\mu_B$) and V MM$_V$ ($\mu_B$), respectively.**

| V Sites | $E_{tot}$ (eV) | $E_{ad}$ (eV) | Distance (Å) | | | | Δρ (e) | MM ($\mu_B$) | MM$_V$ ($\mu_B$) |
|---|---|---|---|---|---|---|---|---|---|
| | | | $d_1$ | $d_2$ | $d_3$ | $d_4$ | | | |
| $H_2$ | -687.05 | 2.18 | 2.21 | 2.21 | 2.21 | 2.21 | 0.41 | 2.84 | 2.59 |
| $H_3$ | -687.39 | 2.52 | 2.15 | 2.15 | 2.15 | 2.15 | 0.48 | 2.69 | 2.86 |
| $H_1$ | -686.69 | 1.82 | 2.31 | 2.31 | 2.32 | 2.32 | 0.43 | 3.87 | 3.28 |
| $B_2$ | -686.62 | 1.75 | 2.16 | 2.16 | 2.17 | 2.17 | 0.38 | 2.97 | 2.99 |

There is obvious charge transfer between V atom and the graphene. For the $H_3$ site,

corresponding the biggest $E_{ad}$, has the biggest charge transfer 0.48 $e$ electron. The V atom has 2.96 $\mu_B$ MM, the whole system has 2.69 $\mu_B$ MM. Each tetragonal carbon atom has -0.01 $\mu_B$ MM, while octagonal carbon atoms have -0.01 and 0.01 $\mu_B$ MM, respectively, shown in Figure 9 (a). For $H_2$ site, 0.41 $e$ electron is transferred from V atom to graphene. And V atom has 2.59 $\mu_B$ MM, the whole system has 2.84 $\mu_B$ MM. m

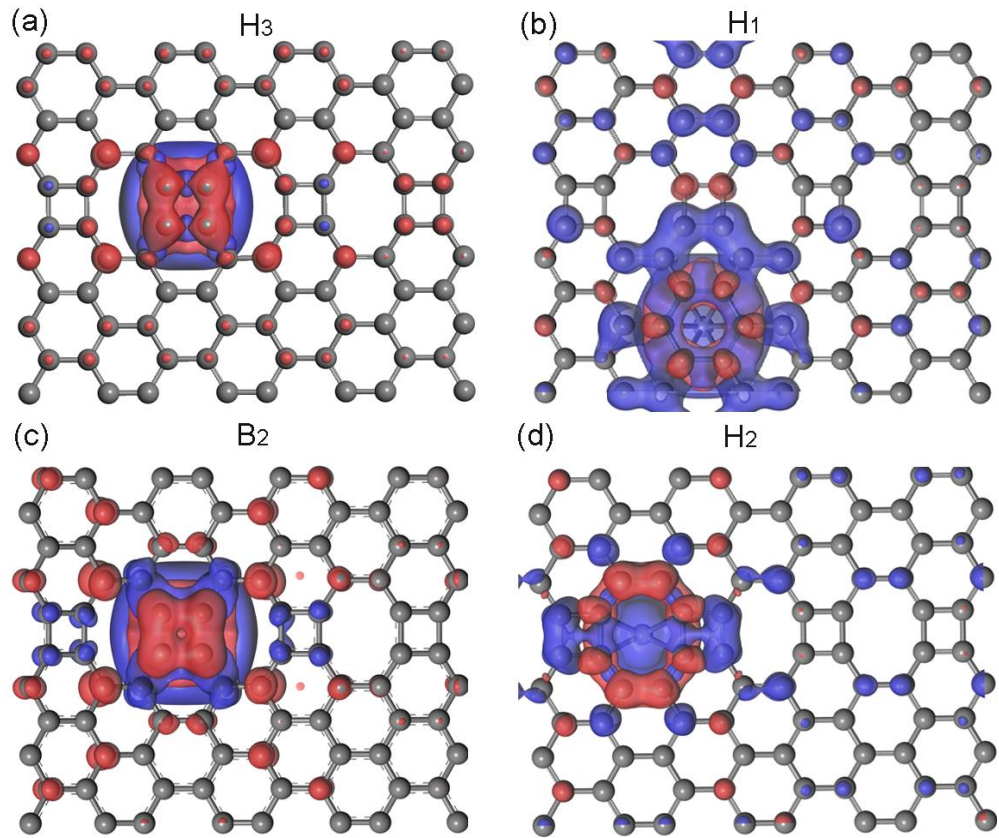

Figure 9. The spin density of single V atom adsorption on the graphene with LD. (a-d) show different adsorption sites, (a) $H_3$, (b) $H_1$, (c) $B_2$ and (d) $H_2$ sites, respectively. The isovalue is set 0.003 e/Å$^3$.

The carbon bonded with carbon atoms has 0.01 $\mu_B$ MM. Other carbon atoms of the octagon have -0.01 $\mu_B$ MM. Carbon atoms of the square have 0.03 and 0.01 $\mu_B$ MM, respectively, shown in Figure 9 (d). While for $H_1$ site, there is 0.41 $e$ electron transfer between V atom and graphene. And corresponding V atom has 3.28 $\mu_B$ MM, and V-LD has 3.87 $\mu_B$ MM, which is consistent with previous results. [45] Each carbon atoms connected with V atom has 0.03 $\mu_B$ MM, shown in Figure 9 (b). For $B_1$ site, V loses 0.38 *e* electron. V atom has 2.99 $\mu_B$ MM, and each carbon atoms connected with V atom has -0.01 $\mu_B$ MM. More details will be found in Figure 9 (c).

As discussed above, the magnetic moment also mainly localizes at adsorption sites. V atom prefers to stay at $H_3$ site, and two V atoms intend to form an "extended metallic wire". The structural, magnetic and electronic properties of "metallic wire" is still unknown. In the following part, the geometry, magnetic and electronic properties are investigated. Based on two V atoms couple method, there is two kinds of magnetic configuration. FM and AFM configurations are calculated shown in Figure 10 (a) and (b), respectively. $E_{FM}$ is -727.04 eV, while $E_{AFM}$ is -726.92 eV. So $\Delta E$ is 0.12 eV, which implies it FM ground state. For AFM configuration, one V atom has 2.89 $\mu_B$ MM, while another V atom has -2.89 $\mu_B$ MM. Each carbon atoms have -0.01$\mu_B$ MM (V atom with 2.89 $\mu_B$), 0.01 $\mu_B$ MM (V atom with -2.89 $\mu_B$) of square, respectively. For lower energy of FM configuration, both two V atoms have 2.89 $\mu_B$ MM. The corresponding $E_{ad}$ is 4.96 eV, which is consistent with $H_3$ site. The spin-polarized band structure and density of state at FM state is also calculated, shown in Figure 10 (c), (d), respectively. The spin-α electron is conductive, while spin-β electron is an insulator with band gap of 0.20 eV. So V-LD is half-metal. Besides, there is Dirac cone composed of spin-α electrons, whose position can be modulated by strains. The half-metallicity is confirmed by analysis of the density of states, and the states near the Fermi-level mainly come from the contribution of V atom and defective carbon atoms.

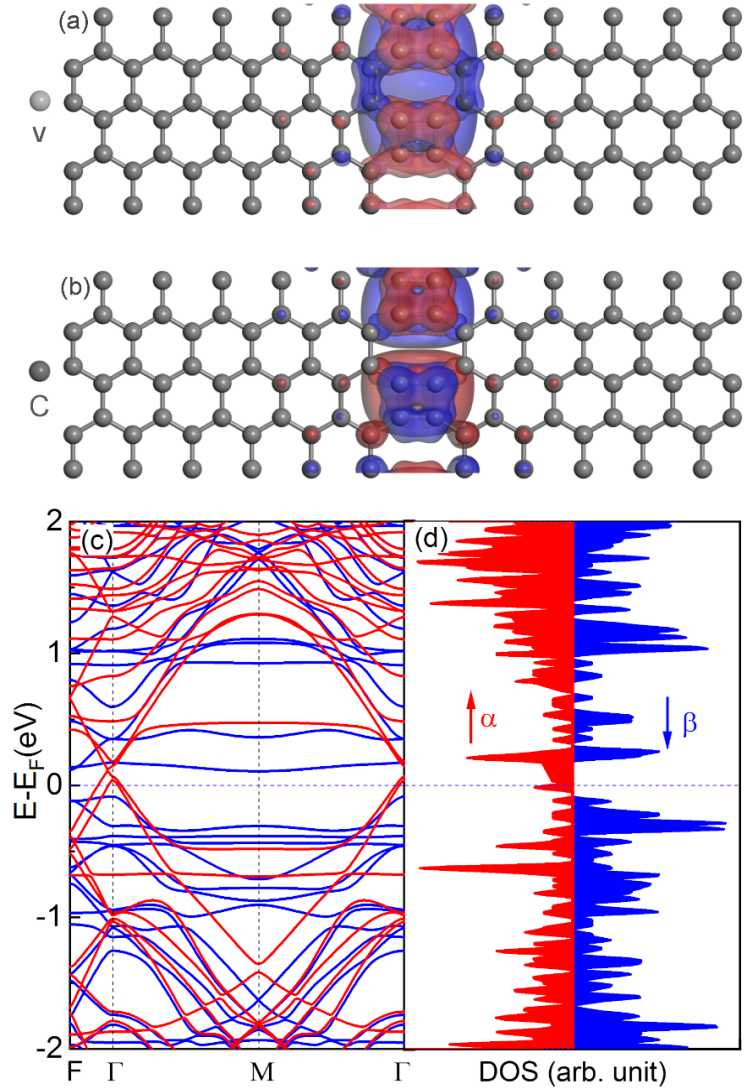

Figure 10. (a-d) Spin density, spin-polarized band structure and PDOS of V-LD. Spin density of (a) FM and (b) AFM configuration of V-LD. (c) Band structure and (d) DOS of V-LD at FM ground state. The blue and red lines represent spin-α and β electrons, respectively. The gray and black present vanadium and carbon atom, respectively. Here, the isovalue is set 0.003 e/Å$^3$.

## 3.3 The Biaxial Strains Tune Electronic Structure Properties of TM-LD

The strain is also usually used as an effective method to modulate electronic properties of low dimensional materials.[19] So the effect of strain on the magnetic and electronic properties of Fe-LD are also investigated. We systematically investigate effect of the enlarged strains. The spin-polarized band structures of Fe-LDs under +2% and +4% strains along $\vec{a}$ direction are calculated, and shown in Figure S5 (a) and (b), respectively, provided in Supporting Information. With enlarged strain of +2%, Fe-LD still presents half-metallic properties. And there is a Dirac cone lying 0.04 eV above Fermi-level for spin-α electrons, while it presents semiconductor with band gap of 0.14 eV for spin-β electrons. As the strain is enlarged to +4%, Fe-LD presents semimetal with Dirac cone for spin-α electrons, while it shows semiconductive properties for spin-β electrons with band gap of 0.13 eV, shown in Figure S5 (b). So smaller enlarged strains could effectively tune Fe-LD spin-polarized metal into half-metal, and even degenerate original Dirac cones composed of spin-α and spin-β electrons.

As discussed above, the strains could effectively tune the position of Dirac cone. So the band structures of Mn-LDs under the enlarged strain are also investigated, shown in Figure S6 in the Supporting information. We mainly considered the enlarged strain of +1%, +2%, +3% and +4%, respectively. And corresponding band structures are shown in Figure S6, (a), (b), (c) and (d), respectively. A Dirac cone of spin-α electron is shifted from the original 0.05 eV above the Fermi-level to the -0.01 eV below the Fermi-level under the 1% enlarged strain. As the enlarged strain is increased to 2%, two

Dirac cone for spin-α electron is continuously shifted -0.015 eV below the Fermi-level. While for other Dirac cones composed of spin-α and spin-β electrons are shifted 0.05 eV above the Fermi-level. As the enlarged strain is increased to 3%, two Dirac cones composed of spin-α electron are shifted downward -0.02 eV below the Fermi-level. And there is a special Dirac cone composed of spin-α and spin-β electrons lying 0.01 eV above the Fermi-level. As the enlarged strain is increased to 4%, two Dirac cones composed of spin-α electrons are shifted down -0.03 eV below the Fermi-level. In a word, the position of Dirac cone can be tuned by the enlarged strains, which means that Mn-LD could be applied in valley and spin electronics.

The stability of the V-LD under the enlarged strains along $\vec{a}$ is also investigated, shown in Figure S8. We can find that V-LD is still at FM ground state, and still show half-metallicity. As the strain increases, the position of Dirac cone is tuned by strains. And the band gap of spin-α electrons monotonously decreases as increase of strains. For the smaller strains (less than 3%), the V-LD still show half-metallic properties. As the strain is increased to 4%, the band edge of spin-α electron is shift upward, even across the Fermi-level. And V-LD changes from half-metal into normal spin-polarized metal. So the stability of half-metallicity of V-LD is well preserved, when the enlarged strain is less than 3%.

**CONCLUSIONS**

In conclusion, we examine the geometry, magnetic, and electronic properties of TM adsorption on the graphene with LD. For Co, Fe, Mn, Ni and V atoms, the most stable configuration is H$_2$ site, which has the largest $E_{ad}$. When single TM (TM = Co, Fe, Mn, V) atom adsorb on the graphene, they introduce localized magnetism around TM and carbon atoms near the TM atom. And corresponding magnetism mainly come from contribution of TM atoms. When two TM atoms adsorb on the graphene, they intend to stay at H$_2$ site, forming a "metallic line", which can work as conductive "metallic wire". All above mentioned metals adsorption on graphene can introduce magnetism and spin-polarization, and they are all at FM ground state, while different metals show different electronic properties. TM-LD (TM = Co, V) is spin-polarized half-metal. TM-LD (TM = Fe, Mn, Ni) are all spin-polarized metal at FM ground state. And uniaxial strains along $\vec{a}$ direction could tune TM-LD into half-metal (TM = Fe, Co, V) or semimetal (TM = Mn) with Dirac cones (composed of one or two kinds of electron's spin). These theoretical findings could open the door to the application in spintronics[9, 57] and valley electrons.[24, 56] This opens up the exciting possibility of the fabrication of carbon based electronic devices with one-dimensional extended LDs that can be used as metallic wire interconnects or elements of device structures.

## ASSOCIATED CONTENT

**Conflict of Interest:** The authors declare no competing financial interest.

**Supporting Information:** This material is available free of charge via the Internet at http://pubs.acs.org.

## AUTHOR INFORMATION

**Corresponding Author**

*E-mail: zyguan@sdu.edu.cn; Tel: +86-0531-88361168; Fax: +86-88361168

ACKNOWLEDGMENT

We thank Doctor Xingxing Li, Haifeng Lv and Professor Wenhui Duan for useful discussion. This project was partially funded by the President foundation of China Academy of Engineering Physics (YZJJLX2016004), the National Key Research and Development Program of China (Grants. No.2016YFB0201203), National Natural Science Foundation of China (Grant No.11904203) and the Fundamental Research Funds of Shandong University (Grant No.*******). The Shanghai Supercomputer Center, National Supercomputing Centers of Guangzhou, and Supercomputer Centers of Tsinghua University should be acknowledged.

**Supporting Information for Publication**

Information on materials, magnetic moment distribution, PDOS of the Co-LD, band structure of Co-LDs under the enlarged strains, PDOS of Fe-LDs, band structures of Mn-LDs under strains, band structures of the V-LDs under the enlarged strains, and band structures of the V-LDs under the enlarged strain.